\documentstyle[preprint,revtex,12pt]{aps}
\draft
\begin{document}
\bibliographystyle{unsrt}
\begin{title}
 Kondo Crossover In The Self-Consistent One-Loop Approximation
\end{title}
\author{Junwu Gan }
\begin{instit}
 Department of Physics, The University of British Columbia,	\\
         6224 Agricultural Road, Vancouver, B.C. Canada V6T 1Z1
\end{instit}

\begin{abstract}
The free energy and magnetization for
the general $SU(N)$ one impurity Kondo model  in the magnetic field, $h$,
are calculated by  extending the previous
$1/N$ expansion technique: the saddle point is determined
self-consistently to the $1/N$ order.  The obtained
universal  field dependent  magnetization $M(h/T_{K})$
by this  simple method is  shown analytically
to be asymptotically exact at  both   $h \ll T_{K}$
and  $h \gg T_{K}$ limits.
For general  ''$f$-electron'' fillings, except half filling,
the $M(h/T_{K})$  curves cross continuously from weak to strong
coupling limit, but overestimate the curvature in the crossover region
for moderate $N$.
  The magnetic Wilson crossover numbers
are calculated for amusement. Our results explicitly verify that
the $1/N$ parameter is non-singular under the adiabatic continuation.
\end{abstract}
%
%


\section{Introduction}

The flowing of an effective interaction
from weak coupling  at high energy to strong coupling  at low energy
is an important and frequently encountered phenomenon in
various physical systems.
 A well known condensed matter
example is the Kondo effect \cite{kondo64}.
Usually, it is
only possible to construct  perturbative
solutions  in the weak and strong coupling limits.
That the Kondo
problem admits  an exact solution provides a useful testbed
for new  ideas and methods.
Among various methods applied to the problem,
 the numerical renormalization group(NRG) \cite{wils75},
Bethe Ansatz \cite{andr83},
and probably Non-Crossing Approximation \cite{bick87-2},
nicely and accurately produce the crossover.
Unfortunately, these methods either  are very  complicated or
heavily rely on numerical calculations.
A simple and elementary method describing
the crossover is desirable and may  give  us new insight.

Recently, motivated by the NRG results
on the two impurity Kondo problem \cite{jones88,jones91}
which claim that there is a line of Fermi liquid fixed points
continuously modified by the RKKY interaction between the two impurity spins,
we have developed an ''Eliashberg equation'' approach
to build the magnetic correlation
between the two impurity spins
nonperturbatively into the ground state \cite{gan-2imp}.
Naturally, we want to test our method for the one impurity Kondo
problem. In this simple case, our approach amounts to  the self-consistent
one-loop approximation.
 For the general $SU(N)$ impurity spin model \cite{coqb69}
with  the orbital degeneracy $N$,
 we expand the free energy in
$1/N$ and determine the saddle point self-consistently
using the free energy including one-loop($1/N$) fluctuation contributions.
We shall see
that $1/N$ is a non-singular
parameter under the adiabatic continuation \cite{coleman86},
at least outside a narrow crossover region.
The effect of high order terms is to smooth out the crossover.
Technically, $1/N$ fluctuations always involve cut-off dependent
contributions. In order to obtain the universal free energy
and magnetization,
all the cut-off dependent terms have to be absorbed into the
Kondo temperature $T_{K}$. In the following,
we first sketch the procedure
then  give the details in the next two sections so that
whoever not interested in details can skip from the end of
introduction directly to the results.

The Kondo problem describes an impurity spin
antiferromagnetically
coupled with strength $J$ to a wide conduction band with
density of states $\rho(\epsilon)$. The Hamiltonian for
the general $SU(N)$  model \cite{coqb69} in the magnetic field is
\begin{equation}
H= \sum_{\vec{k},\sigma} (\epsilon_{\vec{k}} + \sigma h)
c^{\dag}_{\vec{k}\sigma}  c_{\vec{k}\sigma}
-  \frac{J}{N}  \sum_{\vec{k},\vec{k}',\sigma,\sigma'}
(c^{\dag}_{\vec{k}\sigma}f_{\sigma})(f^{\dag}_{\sigma'}c_{\vec{k}'\sigma'})
+ h \sum_{\sigma=-S}^{S} \sigma f^{\dagger}_{\sigma} f_{\sigma}    .
              \label{coqb}
\end{equation}
The impurity spin is represented by $N=2S+1$ localized degenerate levels
partially filled with ''$f$-electrons''.
Their creation and annihilation operators are
subject to the constraint
\begin{equation}
 \hat{n}_{f}=\sum_{\sigma} f^{\dagger}_{\sigma} f_{\sigma} = q_{0} N .
	\label{constraint}
\end{equation}
 We have
set the gyromagnetic ratio and Bohr magneton equal to one
so that  the magnetic field strength $h$ has the energy scale.
For $Ce$, the lower
spin-orbit splitted multiplet usually has $N=6$.
The coefficient $q_{0}$ is treated as a constant of order one \cite{coleman86}
in the expansion and will be given any value at the end of calculation.
We shall present results for $q_{0}=1/2$ and $q_{0}=1/N$.

There are two  physical parameters in the Kondo problem,
the bandwidth $D$ and  the dimensionless coupling constant
$g=J \rho(0)$.
In the scaling regime, $h \ll D$ and $T_{K} \ll D$,  physical
quantities
depend on $D$ and $g$ only through
the Kondo temperature $T_{K}=T_{K}(D, g)$.
If the initial  bare $g \ll 1$,  we can
find $T_{K}$ in the $D/T_{K} \rightarrow \infty$ limit.
This is equivalent to  the ultraviolet renormalization.
The renormalizability of the Kondo problem was stated
long time ago \cite{ande70,abri70} and can be proved
without difficulty.
After absorbing the bare parameters  into $T_{K}$,
physical quantities such as
the magnetization must be a one-variable  function: $M=M(h/T_{K})$,
since $M$ is dimensionless.
Usually,   there could be
many different scaling functions $M(x)$ with $ x=h/T_{K}$, depending
on the  band structures $\rho(\epsilon)$(cut-off schemes).
However, $M(x)$ for the Kondo problem  is universal,
 because changing band structure only adds in irrelevant
perturbations which quickly die out under scaling  if initial
$g \ll 1$ \cite{thank_ian}.
The only possible exception is particle-hole symmetry breaking
perturbation which is marginal and may lead to a modified $M(x)$.
Thus, the obtained scaling solution  for the magnetization
in our calculation is directly comparable with any previous result
up to a proportionality constant between  different
definitions of  the Kondo
temperature \cite{barnes92}.

It has been  known from the phenomenology of dilute alloys \cite{newns80}
that the nature of the strong coupling fixed point of the Kondo problem
is a local resonant level. The two parameters of the resonant
level, its position  $\epsilon_{f}$ and width $\Delta$,
are precisely the saddle point
parameters in the  $1/N$ expansion \cite{read83}.
Including $1/N$ fluctuations,
the free energy  in the magnetic field can be written as
\begin{equation}
F(h, \epsilon_{f}, \Delta, g, D) =
N F_{\rm MF}(h, \epsilon_{f}, \Delta, g, D)
+ F_{1/N}(h, \epsilon_{f}, \Delta, g, D)    ,	\label{f-ener-1}
\end{equation}
where the mean field and
$1/N$ contributions,
$F_{\rm MF}$ and $F_{1/N}$, have no explicit dependence on $N$.
The two parameters $\epsilon_{f}$ and  $\Delta$ are determined by
the stationary condition of the free energy.
To find the Kondo temperature $T_{K}$, we separate out from
the free energy all terms depending on the bare parameters $g$ and $D$,
\begin{equation}
F(h, \epsilon_{f}, \Delta, g, D) =
\widetilde{F}(h, \epsilon_{f}, \Delta, g, D)
+ F_{\rm reg}(h, \epsilon_{f}, \Delta, T_{K})  .
		\label{def-F-reg}
\end{equation}
The regularized free energy, $F_{\rm reg}$,  depends on
$g$ and $D$ only through $T_{K}$.
With a proper definition of $T_{K}$, $\widetilde{F}$ becomes
a constant depending {\em only} on $g$ and $D$,
representing the
correction to the ground state energy.
The thermodynamics is contained in $F_{\rm reg}$
from which we obtain the field dependent magnetization.

The paper is organized as following.
In the next section, we briefly
recapture the large-$N$ approach in the magnetic
field  to define our  notations.
The renormalization procedure  is described in  the third section.
In the fourth section, we present the field dependent magnetization
from  $h \ll T_{K}$ to $h \gg T_{K}$ for several values of $N$.
The magnetic Wilson crossover numbers are  calculated
approximately.
The proof that the  magnetization calculated
from $F_{\rm reg}$ has  the correct
$h \gg T_{K}$ asymptotics
and the integral expressions of  some functions appearing in the
 regularization are included in the appendices for
completeness. To alleviate  cross reference, we list the
frequently occurring symbols together with their defining
 equation numbers in Table~\ref{symbol}.

\section{Large-$N$ Formalism}

Following previous treatments \cite{read83,cole87},
we introduce a Lagrange multiplier $\lambda$
to enforce the constraint~(\ref{constraint}).
By using the fact that the constraint commutes with the Hamiltonian,
we write the partition function in the magnetic field
$h$ as
\begin{eqnarray}
 {\cal Z}  & = & {\rm Tr} \; \delta(\hat{n}_{f}- q_{0} N) \,
\exp[-\beta H  ]
= \int \frac{\beta \, d\lambda}{2\pi}  \; {\rm Tr} \;
\exp\{-\beta[H +i\lambda (\hat{n}_{f} - q_{0} N) ] \}
	\nonumber 		 \\
 &=&  \int \frac{\beta \, d\lambda}{2\pi}
\int {\cal D}[c,\bar{c},f,\bar{f}] \exp \left[ -\int^{\beta}_{0}
d\tau ({\cal L}_{0} + H - i q_{0} N \lambda) \right]			\\
{\cal L}_{0} &=& \sum_{\vec{k},\sigma}  c^{\dag}_{\vec{k}\sigma}
\partial_{\tau} c_{\vec{k}\sigma}
+ \sum_{\sigma}  f^{\dagger}_{\sigma}
 (\partial_{\tau} + i\lambda )f_{\sigma}   .
\end{eqnarray}
After performing
Hubbard-Stratonovich transformation to factorize the Kondo interaction,
we rewrite the partition function as
\begin{eqnarray}
{\cal Z} &=& \int \frac{\beta \, d\lambda}{2\pi}
\int {\cal D}[c,\bar{c},f,\bar{f},Q,\bar{Q}]
\exp\left[ - \int^{\beta}_{0} d \tau
\left( {\cal L}_{0} + {\cal L}'
+ \frac{N |Q|^{2} }{J}   - i q_{0} N  \lambda  \right) \right]	       \\
{\cal L}' &=&  \sum_{\vec{k},\sigma} ( \epsilon_{\vec{k}} + \sigma h)
c^{\dag}_{\vec{k}\sigma} c_{\vec{k}\sigma}
 + \sum_{\vec{k}, \sigma} ( Q c^{\dag}_{\vec{k}\sigma} f_{\sigma}
+\bar{Q} f^{\dag}_{\sigma} c_{\vec{k}\sigma} )
+ h \sum_{\sigma} \sigma f^{\dagger}_{\sigma} f_{\sigma}  .
\end{eqnarray}
The above Lagrangian possesses a U(1) gauge invariance
\begin{equation}
f_{\sigma} \rightarrow f'_{\sigma} = f_{\sigma} \,  e^{i\phi} ,
\hspace{.2in}
Q      \rightarrow    Q' = Q \, e^{-i\phi}   ,
\hspace{.2in}
\lambda \rightarrow \lambda'=\lambda + \frac{d \phi}{d \tau} .
\end{equation}
The redundant  gauge degrees of freedom can be eliminated by
choosing to work in the radial gauge.
Separating the complex field $Q$ into an amplitude and a phase
$ Q = r \, e^{-i\phi}$, the phase $\phi$
can be absorbed into new variables $f'_{\sigma}$ and $\lambda'$:
 $ f'_{\sigma} = f_{\sigma} \, e^{-i\phi}$,
 $ \lambda'=\lambda + d\phi/d\tau$ .
In terms of new variables $r$,  $\lambda'$,
$f'_{\sigma}$ and $\bar{f}'_{\sigma}$,
the partition function can be cast in the form, after
dropping the primes,
\begin{equation}
{\cal Z} = \int {\cal D}[c,\bar{c},f,\bar{f},\lambda,r]  \,
\prod_{\tau} r(\tau) \,
\exp\left[ - \int^{\beta}_{0} d \tau
\left( {\cal L}''(\tau) + \frac{N r^{2} }{J} -i  q_{0} N \lambda
		\right) \right]
  \end{equation}
\begin{equation}
{\cal L}'' = \sum_{\vec{k},\sigma} c^{\dag}_{\vec{k}\sigma}
(\partial_{\tau} + \epsilon_{\vec{k}} + \sigma h ) c_{\vec{k}\sigma}
+\sum_{\sigma=-S}^{S}
f^{\dag}_{\sigma} (\partial_{\tau} + i\lambda + h\sigma ) f_{\sigma}
+ \sum_{\vec{k}\sigma} r \,(  c^{\dag}_{\vec{k}\sigma}f_{\sigma}
+ f^{\dag}_{\sigma} c_{\vec{k}\sigma} )  .
\end{equation}
It is possible to completely gauge away  the $U(1)$ phase $\phi$ because
it does not contain dynamics.
Since the last Lagrangian is bilinear in the Grassman variables
$c_{\vec{k}\sigma}$ and $f_{\sigma}$,
we can  integrate them out to obtain an effective action,
\begin{eqnarray}
{\cal Z} &=& {\cal Z}_{0} \, \int {\cal D}[\lambda,r]  \,
\exp[- S_{\rm eff}(\lambda,r)
+ \delta(0) \int^{\beta}_{0} d\tau \ln r(\tau)]               \\
S_{\rm eff} &=& - \sum_{\sigma} {\rm Tr} \,
\ln \left[ \partial_{\tau} + i\lambda + h\sigma + r G_{0}(\tau) r \right]
+ N \int^{\beta}_{0} d \tau
\left(\frac{r^{2}}{J} - i q_{0} \lambda \right)   ,
\end{eqnarray}
where $\delta(0) = (1/\beta) \sum_{\nu_{n}} 1 $ with
$\nu_{n}=2 \pi n /\beta$,  and
\begin{equation}
G_{0}(\tau) = - \sum_{\vec{k}}
\frac{1}{ \partial_{\tau} + \epsilon_{\vec{k}} }  .	\label{g0}
\end{equation}
${\cal Z}_{0}$ is the partition
function of the non-interacting Fermi sea.

The integration over the two real variables $\lambda$ and $r$
can be expanded around a saddle point
\begin{equation}
i \lambda= \epsilon_{f} +  i \widetilde{\lambda},
\hspace{.4in} r=r_{0} + \widetilde{r} .		\label{sad-expan}
\end{equation}
Retaining only quadratic terms in  $\widetilde{\lambda}$
and $\widetilde{r}$ in the expansion, the partition function
becomes, after dropping the tilde sign,
\begin{equation}
\frac{{\cal Z}}{{\cal Z}_{0}} = e^{-S_{\rm eff}(\epsilon_{f},r_{0})}
\int \prod_{\nu_{n}} d\lambda(\nu_{n}) \, dr(\nu_{n}) \;
\exp \left[ - S_{\rm eff}^{(2)}+ \sum_{\nu_{n}} \ln r_{0} \right]
			\label{s-eff-1}
\end{equation}
\begin{equation}
S_{\rm eff}^{(2)} =
\frac{N}{2} \sum_{\nu_{n}}
(\lambda(-\nu_{n}), r(-\nu_{n}))
\left(	\begin{array}{cc}
 \rho(0) r_{0}^{2}  \Gamma_{\lambda}(\nu_{n}) &
i \rho(0) r_{0}  \Gamma_{\lambda r}(\nu_{n})	\\
i \rho(0) r_{0}  \Gamma_{\lambda r}(\nu_{n})	&
\rho(0) \Gamma_{ r}(\nu_{n})
\end{array}			\right)
\left(		\begin{array}{c}
 \lambda(\nu_{n}) \\ r(\nu_{n}) 	\end{array} \right)	.
			\label{s2-eff}
\end{equation}
The zero temperature expressions of the matrix elements $\Gamma$'s appearing
in  $ S_{\rm eff}^{(2)} $ have been given
by Read \cite{read83}. Their extension to include magnetic field
is straightforward.
Here we have  pulled out explicitly some prefactors for later convenience.
 \begin{eqnarray}
\Gamma_{\lambda}(\nu_{n}) &=& \frac{1}{N} \sum_{\sigma}
\frac{ 1 }{  |\nu_{n}| (|\nu_{n}| + 2 \Delta ) }
\ln \left[ \frac{\epsilon_{f\sigma}^{2} + (|\nu_{n}| + \Delta)^{2} }
{\epsilon_{f\sigma}^{2} + \Delta^{2} } \right]
	\label{Gam-l}	\\
\Gamma_{\lambda r}(\nu_{n}) &=& -\frac{2}{N |\nu_{n}|} \sum_{\sigma} \left[
\tan^{-1} \left( \frac{\epsilon_{f\sigma} }{|\nu_{n}| + \Delta} \right)
-\tan^{-1} \left( \frac{\epsilon_{f\sigma} }{\Delta} \right) \right]
	\label{Gam-lr}	\\
\Gamma_{r}(\nu_{n}) &=& \frac{1}{N} \sum_{\sigma}
\left\{ \ln \left[ \frac{\epsilon_{f\sigma}^{2} + (|\nu_{n}| + \Delta)^{2} }
{ (T_{K}^{(0)})^{2} } \right] + \frac{2\Delta}{|\nu_{n}|}
\ln \left[ \frac{\epsilon_{f\sigma}^{2} + (|\nu_{n}| + \Delta)^{2} }
{\epsilon_{f\sigma}^{2} + \Delta^{2} } \right]	\right\}	,
			\label{Gam-r}
\end{eqnarray}
where we  have defined the mean field  Kondo temperature,
\begin{equation}
T_{K}^{(0)} = D \exp \left(-\frac{1}{g} \right), \hspace{.3in}
	 g=J \rho(0) , 		\label{Tk0}
\end{equation}
 and the convenient notations,
\begin{equation}
\epsilon_{f\sigma} =\epsilon_{f} +\sigma h,
\hspace{.4in} \Delta= \pi \rho(0) r_{0}^{2} .   \label{def-Delta}
\end{equation}
The contributions to the free energy~(\ref{f-ener-1})
are given by
\begin{eqnarray}
F_{\rm MF} &=& \frac{1}{N} \sum_{\sigma}
\left\{ \frac{\epsilon_{f\sigma}}{\pi}
\tan^{-1}\left(\frac{ \epsilon_{f\sigma}}{\Delta} \right)
+ \frac{\Delta}{2\pi}
\ln \left[ \frac{\epsilon_{f\sigma}^{2}  + \Delta^{2} }
{ (T_{K}^{(0)})^{2} } \right]  \right\} -\frac{\Delta}{\pi}
+ \left(\frac{1}{2} - q_{0} \right) \epsilon_{f}	\label{F-mf}					\\
F_{1/N} &=& \frac{1}{2\beta} \sum_{\nu_{n}}
\ln [ \Gamma_{\lambda}(\nu_{n}) \Gamma_{r}(\nu_{n})
 + \Gamma_{\lambda r}^{2}(\nu_{n}) ] + {\rm const.}
\label{F1}
\end{eqnarray}
In the free energy $F_{1/N}$,
we note that the prefactors in the front of $\Gamma$'s in (\ref{s2-eff})
exactly cancel the    contribution
$ \sum_{\nu_{n}} \ln r_{0} $  of (\ref{s-eff-1}),
originating from the Jacobian of transforming to the radial gauge.

\section{Renormalization}

To calculate zero temperature quantities, we can simply replace
the discrete Matsubara frequency sum by  an integration
\begin{equation}
F_{1/N} = \frac{1}{2\pi} \int^{\infty}_{0} d\nu
\ln ( \Gamma_{\lambda} \Gamma_{r} + \Gamma_{\lambda r}^{2} ) ,
 \hspace{.3in}  \frac{1}{\beta}\sum_{\nu_{n}}
 \rightarrow \int^{\infty}_{-\infty} \frac{d\nu}{2\pi},
\hspace{.3in}
|\nu_{n}| \rightarrow \nu	.	\label{sum-int}
\end{equation}
The upper integration limit is actually cut off by the conduction electron
bandwidth $D$. One can see this from the
 approximation we made in deriving the
mean field free energy and $1/N$ fluctuation matrix element $\Gamma$'s,
\[
 \sum_{\vec{k}} \frac{1}{i\omega_{n} - \epsilon_{\vec{k}} }=
\rho(0) \int^{D}_{-D} \frac{d\epsilon}{i\omega_{n}-\epsilon}
= -i 2  \rho(0) \tan^{-1}\left(\frac{D}{\omega_{n}}\right)
\simeq - i \pi \rho(0)  \; {\rm sgn}\omega_{n}  \;
\theta(D-|\omega_{n}|).
				\]
Obviously,  $F_{1/N}$ of
(\ref{F1}) contains contributions linear in $D$
which become divergent in the $D \rightarrow \infty$ limit.
A little investigation shows that the sub-leading
divergent terms of $F_{1/N}$ have the form of
 $\ln\ln D$.

To separate  out the cutoff dependent
 terms of $F_{1/N}$ which diverge as $D \rightarrow \infty$,
we consider the $\nu \rightarrow \infty$  asymptotic behavior
of the integrand,
\begin{equation}
\Gamma(\nu) =  \Gamma_{\lambda} \Gamma_{r} + \Gamma_{\lambda r}^{2}
= \frac{1}{\nu^{2}}
\left[ \Gamma_{1}(\ln \nu) + \frac{2}{\nu} \Gamma_{2}(\ln \nu)
+{\cal O}(\nu^{-2}) \right] .		\label{Gam-exp}
\end{equation}
 The two functions
$\Gamma_{1}$ and $\Gamma_{2}$  only depend on   $\ln \nu$
and have the following simple forms
\begin{eqnarray}
\Gamma_{1}(\ln \nu) &=& 4 \left[ \ln^{2}\frac{\nu}{T_{K}}
- \pi \eta_{2} \, \ln\frac{\nu}{T_{K}}
 +  \pi^{2} \left( \frac{1}{2} - q_{0} - \eta_{1}
\right)^{2} \right]  , 			\label{Gam-1}	\\
\Gamma_{2}(\ln \nu) &=& 4 \left[ \Delta (1-\pi \eta_{2})
\ln\frac{\nu}{T_{K}} - \pi \eta_{2}  \Delta
 \left(\frac{1}{2} - \pi \eta_{2} \right)
- \pi \epsilon_{f} \left( \frac{1}{2}-q_{0}-\eta_{1} \right)
 \right]  ,    \label{Gam-2}
\end{eqnarray}
where we have introduced following two short hand notations,
\begin{eqnarray}
\eta_{1} &=&  \frac{\partial F_{\rm MF}(\epsilon_{f}, \Delta) }
{\partial \epsilon_{f} }
= \frac{1}{2} - q_{0} - \frac{1}{\pi N} \sum_{\sigma}
\tan^{-1}\left(\frac{ \epsilon_{f\sigma}}{\Delta} \right)  ,
	\label{def-eta1}	\\
\eta_{2} &=&  \frac{\partial F_{\rm MF}(\epsilon_{f}, \Delta) }
{\partial \Delta}
  = \frac{1}{N \pi }\sum_{\sigma}
\ln \left( \frac{\sqrt{\epsilon_{f\sigma}^{2}
+ \Delta^{2}}}{T_{K}^{(0)}} \right) 		 .	\label{def-eta2}
\end{eqnarray}
They both are independent of frequency $\nu$.
The $1/N$ fluctuation free energy is regularized as following,
\begin{eqnarray}
F_{1/N} &=& \int^{\infty}_{0} \frac{d\nu}{2\pi}
\left\{ \ln \Gamma(\nu) -  \left[ \ln \Gamma_{1}(\ln \nu) +
\frac{ 2 \,  \Gamma_{2}(\ln \nu)}{ \nu \,  \Gamma_{1}(\ln \nu)} \right]
 \theta(\nu-\nu_{0}) \right\}    \nonumber		\\
 &  & + \int^{D}_{\nu_{0}}  \frac{d\nu}{2\pi} \ln \Gamma_{1}(\ln \nu)
+ \int^{\ln (D/T_{K}) }_{\ln (\nu_{0}/T_{K}) } \frac{dx}{\pi}
\frac{\Gamma_{2}(x)}{ \Gamma_{1}(x)} + {\rm const} .
		\label{F1-sep}
\end{eqnarray}
Since the first integral is convergent, we  have extended the upper
integration limit
to infinity. Note that $\nu_{0}$ is not a parameter of the theory.
$F_{1/N}$ is independent of $\nu_{0}$. We shall choose it
for computational convenience. Actually,
it provides a useful consistency
check for the numerical  calculation.
The cut-off dependence is then separated out from the last
two integrals of (\ref{F1-sep}),
\begin{eqnarray}
\frac{1}{2\pi} \int^{D}_{\nu_{0}}  d\nu \ln \Gamma_{1}(\ln \nu)
&=& D \, \Lambda_{1}(D, \eta_{1}, \eta_{2})
 - \nu_{0}  \, \Lambda_{1}(\nu_{0}, \eta_{1}, \eta_{2} )
			\label{L1}		\\
\frac{1}{\pi} \int^{\ln (D/T_{K}) }_{\ln (\nu_{0}/T_{K}) } dx
\frac{\Gamma_{2}(x)}{ \Gamma_{1}(x)}
&=& \Lambda_{2}(D, \eta_{1}, \eta_{2}, \epsilon_{f}, \Delta)
 -  \Lambda_{2}(\nu_{0}, \eta_{1}, \eta_{2}, \epsilon_{f}, \Delta)  .
			\label{L1L2}
\end{eqnarray}
The so-defined  two functions $\Lambda_{1}$ and $\Lambda_{2}$
 are given in the appendix.

To treat the cut-off dependent terms
 $ D \,  \Lambda_{1}(D, \eta_{1}, \eta_{2})  $
and $ \Lambda_{2}(D, \eta_{1}, \eta_{2}, \epsilon_{f}, \Delta)$,
we first obtain explicitly
\begin{equation}
\Lambda_{2}(D, \eta_{1}, \eta_{2}, \epsilon_{f}, \Delta) =
 \frac{\Delta}{\pi}  \ln\ln \frac{D}{T_{K}}
  -  \eta_{2}	\;  \Delta \ln\ln \frac{D}{T_{K}} ,
		\label{L2}
\end{equation}
where we have neglected terms which vanish as $D \rightarrow \infty$.
Using the fact  that $\eta_{1}$ and $\eta_{2}$ are
the derivatives of the mean field free energy,
we can show  that
$\Lambda_{1}$ and the second  term
of (\ref{L2})  can be renormalized away
from the saddle point equations if we let
the saddle point parameters $\epsilon_{f}$ and $\Delta$ acquire
following $1/N$ corrections,
\begin{eqnarray}
\widetilde{ \epsilon}_{f} &=& \epsilon_{f} +
 \frac{D}{N} \frac{\partial }{\partial \eta_{1} }
  \Lambda_{1}(\eta_{1}^{*}, \eta_{2}^{*})   \\
\widetilde{ \Delta} &=& \Delta  - \frac{\Delta}{N}  \ln\ln \frac{D}{T_{K}}
  +  \frac{D}{N}   \frac{\partial }{\partial \eta_{2} }
  \Lambda_{1}(\eta_{1}^{*}, \eta_{2}^{*})
 \end{eqnarray}
where $\eta_{1}^{*}$ and $\eta_{2}^{*}$ are the
 values at the point of the  saddle point solution,
$\epsilon_{f} = \epsilon_{f}^{*}$ and $ \Delta=\Delta^{*}$.
When we rewrite the mean field free energy in terms of the renormalized
saddle point parameters $\widetilde{ \epsilon}_{f}$ and $\widetilde{ \Delta}$,
we have to include the difference
$F_{\rm MF}(\epsilon_{f},\Delta)
- F_{\rm MF}(\widetilde{\epsilon}_{f},\widetilde{\Delta})$
into  the cut-off dependent part of the free energy $\widetilde{F}$
introduced in
(\ref{def-F-reg}).  Collecting this term,  (\ref{L2}),
$\Lambda_{1}(D, \eta_{1}, \eta_{2}) $,
and a term coming from replacing $T_{K}^{(0)}$
by $T_{K}$ in $F_{\rm MF}$, the total cut-off dependent
part of the free energy is
\begin{equation}
\widetilde{F} = - N \frac{\Delta}{\pi}
 \ln \frac{T_{K}^{(0)}}{T_{K}}
+ \frac{\Delta}{\pi} \ln \ln \frac{D}{T_{K}}
+ D \left[ \Lambda_{1}(\eta_{1}, \eta_{2})
- \frac{\partial \Lambda_{1}(\eta_{1}^{*}, \eta_{2}^{*}) }{\partial \eta_{1}}
\eta_{1}
- \frac{\partial \Lambda_{1}(\eta_{1}^{*}, \eta_{2}^{*}) }{\partial \eta_{2}}
\eta_{2}	 \right]		,	\label{F-til-exp}
\end{equation}
 Note that the last term
is a constant, to the order
${\cal O}(\eta_{1}) \sim {\cal O}(\eta_{2})$.
The first two terms cancel out if we define
\begin{equation}
T_{K}= T_{K}^{(0)}  \left( \ln\frac{D}{T_{K}} \right)^{-1/N}
=D \left( \ln\frac{D}{T_{K}} \right)^{-1/N}
\exp \left(- \frac{1}{g} \right) .
		\label{kondo-T}
\end{equation}
In the spirit of order by order renormalization,
 we replace $\epsilon_{f}$,
$\Delta$ and $T_{K}^{(0)}$ appearing in $F_{1/N}$
by $\widetilde{\epsilon}_{f}$,
$\widetilde{\Delta}$ and $T_{K}$ respectively.
This gives us the regularized free energy as a function of $h$,
$\widetilde{\epsilon}_{f}$,
$\widetilde{\Delta}$ and $T_{K}$ only.
Note that our expression for the Kondo temperature is consistent
with the well known expression $T_{K}=D g^{1/N} \exp(-1/g)$
up to  ${\cal O}(1/N)$.

Actually, one can simply expand $\Lambda_{1}(\eta_{1},\eta_{2})$
in $1/N$ by using the fact $\eta_{1} \sim \eta_{2} \sim {\cal O}(1/N)$,
a consequence of the saddle point equations. We immediately see that
the only ${\cal O}(1)$ contribution of
$\Lambda_{1}(\eta_{1},\eta_{2})$ to the free energy  is a constant.
This constant is the correction to the ground state energy
and has no effect on the physical quantities.
Higher order terms in the expansion of $\Lambda_{1}(\eta_{1},\eta_{2})$
can be neglected in the order by order renormalization.
The second term of  (\ref{L2}) is also dropped since it is
of order ${\cal O}(1/N)$.
After we renormalize away the first term of  (\ref{L2})
by defining the $1/N$ corrected Kondo temperature $T_{K}$ via
 (\ref{kondo-T}) and
 replace the mean field Kondo temperature $T_{K}^{(0)}$  in
$\Gamma_{r}$ by $T_{K}$,
the resulting regularized free energy is then only
a function of $\epsilon_{f}$, $\Delta$, $h$ and $T_{K}$.
All these are due to the fact that the free energy is stationary
with respect to $\epsilon_{f}$ and $\Delta$.  A ${\cal O}(1/N)$ shift
of these parameters does not induce any change in the free energy to
 the  order ${\cal O}(N) + {\cal O}(1)$.

After completing the renormalization,   the
 universal free energy is,  from (\ref{def-F-reg}) and
(\ref{F1-sep})-(\ref{L1L2}),
\begin{eqnarray}
F_{\rm reg} &=&   \sum_{\sigma}  \frac{\epsilon_{f\sigma}}{\pi}
\tan^{-1}\left(\frac{ \epsilon_{f\sigma}}{\Delta} \right)
- \frac{N \Delta}{\pi}  \left[ 1  -  \frac{1}{N}\sum_{\sigma}
\ln \left( \frac{\sqrt{\epsilon_{f\sigma}^{2}
+ \Delta^{2}}}{T_{K}} \right) 		\right]   \nonumber  \\
 & & \hspace{.5in}
 + N \left(\frac{1}{2} - q_{0} \right) \epsilon_{f}
+ F_{1/N}^{\rm reg}		\label{Freg-fin}	\\
 F_{1/N}^{\rm reg}	&=&
- \nu_{0}   \, \Lambda_{1}(\nu_{0}, \eta_{1}, \eta_{2})
 -  \Lambda_{2}(\nu_{0}, \eta_{1}, \eta_{2}, \epsilon_{f}, \Delta) 		\nonumber
 \\
 & + &  \int^{\infty}_{0} \frac{d\nu}{2\pi}
\left\{ \ln \Gamma(\nu) -  \left[ \ln \Gamma_{1}(\ln\nu) +
\frac{2 \, \Gamma_{2}(\ln\nu) }{\nu \, \Gamma_{1}(\ln\nu)} \right]
 \theta(\nu-\nu_{0}) \right\}   .			\label{F-final}
\end{eqnarray}
The parameters $\eta_{1}$ and $\eta_{2}$
only depend on $\epsilon_{f}$, $\Delta$.
Inside $\eta_{2}$  and
$\Gamma_{r}$,
$T_{K}^{(0)}$ is replaced by $T_{K}$.

The saddle
point parameters, $\epsilon_{f}$ and $\Delta$, are determined
 by solving  the following two  saddle point equations,
\begin{eqnarray}
 & &  \frac{1}{N} \frac{\partial}{\partial \epsilon_{f}}
F_{\rm reg}(h, \epsilon_{f}, \Delta, T_{K}) =
 \frac{1}{2} - q_{0} - \frac{1}{\pi N} \sum_{\sigma}
\tan^{-1}\left(\frac{ \epsilon_{f\sigma}}{\Delta} \right)
+ \frac{1}{N} \frac{\partial}{\partial \epsilon_{f}}  F_{1/N}^{\rm reg}
	=0			\label{sad1-expre} \\
& &  \frac{1}{N} \frac{\partial}{\partial \Delta}
F_{\rm reg}(h, \epsilon_{f}, \Delta, T_{K}) =
 \frac{1}{\pi }\sum_{\sigma}
\ln \left( \frac{\sqrt{\epsilon_{f\sigma}^{2}
+ \Delta^{2}}}{T_{K}} \right)
+  \frac{\partial}{\partial \Delta}  F_{1/N}^{\rm reg}  =0
			\label{sad2-expre}    .
\end{eqnarray}
Substituting the solution
$\epsilon_{f}=\epsilon_{f}^{*}(h/T_{K})$ and
 $\Delta=\Delta^{*}(h/T_{K})$
back into $F_{\rm reg}$, we obtain the scaling form of the free
energy depending only on $h/T_{K}$, up to an additive constant.
The magnetization is
\begin{equation}
M(h/T_{K})  =  - \frac{\partial}{\partial h}
F_{\rm reg}(h, \epsilon_{f}^{*}, \Delta^{*}, T_{K})
 = \frac{1}{\pi}  \sum_{\sigma} \, \sigma \,
\tan^{-1}\left(\frac{ \epsilon_{f\sigma}}{\Delta} \right)
-  \frac{\partial}{\partial h}  F_{1/N}^{\rm reg}	 .
		 \label{m-h-expre}
\end{equation}
The  one-dimensional integration  in
the regularized $1/N$ free energy
and its derivatives,
as well as  solving the two coupled equations
(\ref{sad1-expre}) and (\ref{sad2-expre}),
 are carried out numerically.

We emphasize that the obtained magnetization
is {\em not} a  $1/N$
perturbative result if we solve the
equations~(\ref{sad1-expre})
and (\ref{sad2-expre}) self-consistently, {\em i.e.} not by
expanding $\epsilon_{f}^{*}$ and $\Delta^{*}$ in $1/N$.
The fact that we only carried out perturbative ultraviolet
renormalization only  implies that the Kondo temperature defined by
(\ref{kondo-T}) is perturbatively accurate to
the $1/N$ order.
In other words, our result for  $F_{\rm reg}$
or $M(h/T_{K})$
is perturbative
at high energy but not necessarily perturbative at low energy,
depending on how we solve the saddle point equations.
As we can see, the same renormalization
procedure can be carried
out for every physical quantity and
their calculation is a straightforward exercise.

\section{Results}

The solution of the saddle point equations, $\epsilon_{f}^{*}(h/T_{K})$
and $\Delta^{*}(h/T_{K})$, for $q_{0}=1/6$, $N=6$ is shown
in Fig.~{\ref{fig-sadd} as an example.
Generally for $q_{0} \neq 1/2$,
there are more than one solution in the weak coupling regime
for a given value of $h/T_{K}$.
Certainly, the criterion is to choose  one with  the lowest
energy. However, since we know the asymptotics
at both weak and strong coupling limits,
we can follow the solutions continuously by   varying the
magnetic field slightly each time.
For $q_{0}=1/6$ and $N=6$ as an example,
there are solutions other
than that shown in  Fig.~{\ref{fig-sadd} for $h/T_{K} > 0.52$ and give
magnetizations much closer to
Hewson and Rasul's exact results \cite{hews83}
 in near crossover region compared with the results
shown in Fig.~\ref{fig-coqb}.
But, if we follow these solutions
 to high magnetic field, they do not have  the correct
asymptotics.

The field dependent magnetizations $M(h/T_{K})$ for $q_{0}=1/2$
and various  values of $N$ are shown in Fig.~\ref{fig-qhalf}.
Note that each curve has a window in the crossover region where no solution
is found by the present method. This happens only for $q_{0}=1/2$.
The reason is following. We try to describe the strong coupling
fixed point by a resonant level.
The particle-hole symmetry presented in the $q_{0}=1/2$ case
ties the position of the resonant level at the Fermi surface,
$\epsilon_{f}^{*}=0$, in the strong coupling regime.
Certainly,  the nature of the weak coupling is no longer
a resonant level, thus $\epsilon_{f}^{*} \neq 0$.
A discontinuity must  occur at some value
of $\epsilon_{f}^{*} $  with increasing  magnetic field $h$,
preventing continuous
crossover from one side to the other.
Nevertheless, the window quickly narrows with increasing $N$.
For $N=8$, the solid line of Fig.~\ref{fig-qhalf},
the window narrows to $0.45 < h/T_{K} < 0.55$.
The indication is that probably we need infinite
order of terms  in $1/N$ to close the window and
to  obtain completely  smooth crossover.
The more terms we put in, the better the quality is
 in the crossover region.
Similar features can also be seen  for general values of $q_{0}$.
 In Fig.~\ref{fig-coqb}, we show
the magnetizations for $q_{0}=1/N$,
the ''realistic'' situation. Also shown are Hewson and Rasul's Bethe
Ansatz results  \cite{hews83,hews83-1} for $N=6, \, 8$.
Although the lines can  cross
continuously from one side to the other, they obviously
overestimate the curvature in the crossover region.
With increasing $N$, the curvature is reduced.

For amusement, we calculate the magnetic Wilson crossover numbers
for the Coqblin-Schrieffer model \cite{coqb69}, $q_{0}=1/N$, although the
calculation  can be done
for other values of $q_{0}$.
 The ambiguity in relating
$T_{K}$ from different cutoff schemes can be eliminated
by imposing  the condition of a vanishing $\ln^{-2}(h/T_{K})$ term in
the  $h/T_{K} \gg 1$ expansion of $M(h/T_{K})$.
The weak coupling scaling form for the magnetization in terms of $T_{K}$
is well known \cite{andr83},
\begin{equation}
\frac{M}{M_{0}} = 1 - \frac{1}{2 \ln \frac{h}{T_{K}} }
- \frac{ \ln \ln  \frac{h}{T_{K}} }{ 2 N \ln^{2}  \frac{h}{T_{K}} }
+ \frac{  \ln 2 }{ N \ln^{2}  \frac{h}{T_{K}} } + \cdots  ,
	\hspace{.3in}  h/T_{K} \gg 1 .
			\label{m-high-h}
\end{equation}
The last term of (\ref{m-high-h}) can be  removed by changing to a new
 energy scale
\begin{equation}
T_{h} = 2^{-2/N} \, T_{K} \simeq T_{K} / \left(1 + \frac{2 \ln 2}{N} \right) .
			\label{def-Th}
\end{equation}
Although we only explicitly prove the first log term of
(\ref{m-high-h}) in the appendix, we expect that
our result~(\ref{m-h-expre})
will precisely produce
 all three log terms of (\ref{m-high-h}), since all $1/N$ order
contributions to the free energy are included in the present approach.
Another  direct way to see this is following.
Given the second term of (\ref{m-high-h}),
the last  two
terms of (\ref{m-high-h}) are determined by
the second term of the weak coupling beta function \cite{ande70},
\begin{equation}
\beta(g) = \frac{ d g}{d \ln D} = - g^{2} + \frac{g^{3}}{N} ,
		\label{beta-f}
\end{equation}
Our expression for the
Kondo temperature (\ref{kondo-T}) gives  exactly the same beta function.
The correct asymptotic form (\ref{m-high-h})
 allows unambiguous
determination of the energy scale $T_{h}$ in the present approach.
In terms of the unique energy scale $T_{h}$, the
coefficient $\alpha'$ in the strong coupling asymptotic
form of the magnetization
\begin{equation}
\frac{M}{M_{0}} = \alpha \frac{h}{T_{K}}
 = \alpha' \frac{h}{T_{h}}  , \hspace{.4in}
		\frac{h}{T_{h}} \ll 1, 		\label{give-cn}
\end{equation}
is just the magnetic Wilson crossover number. From (\ref{give-cn})
and (\ref{def-Th}), we see $\alpha'=\alpha/(1+2\ln 2/N)$.
The slope $\alpha$ will be determined directly
 from $M(h/T_{K})$ curve.
We list the results
for the general $SU(N)$ cases in Table~\ref{c-number}.

In summary, we calculated the universal field
dependent magnetization for the general $SU(N)$ one impurity Kondo model
for various values of $N$ and ''$f$-electron'' fillings.
At both small and high field limits, our results
become asymptotically exact, as shown
analytically in the appendix.
 For other than  half  filling of the ''$f$-electrons'',
 the magnetization curves cross continuously from one side
to the other. In the crossover region,
the bigger  is the $N$, the smoother and the
more accurate is the magnetization. In contrast to
a continuous phase transition,
the crossover involves no divergence.
The other facet
of the story is that one then does need high order terms to
 smooth out the crossover for a given $N$.

It is my pleasure to thank Ian Affleck, Natan Andrei, Piers Coleman,
Eric Sorenson, Eugene Wong for many helpful discussions.
Illuminating discussion with Affleck helped me to understand many
crucial points. I am also grateful to
Alex Hewson. He kindly provided me the Bethe Ansatz results.
This work was supported by NSERC of Canada.

\appendix{Integrals $\Lambda_{1}$ and $\Lambda_{2}$}

For simplicity, we set  $T_{K}=1$ in this section.
 From the definition,
$\Lambda_{2}$  is an integral  of the  type,
\begin{equation}
\Lambda_{2}(D,\eta_{1},\eta_{2},\epsilon_{f},\Delta)
 - \Lambda_{2}(\nu_{0}, \eta_{1},\eta_{2},\epsilon_{f},\Delta)
 = \int^{D}_{\nu_{0}}
\frac{d\nu}{\pi} \frac{ w \ln\nu + v }
{\nu \left( \ln^{2}\nu + a \ln\nu + b \right)} ,
\end{equation}
where $a$, $b$, $w$ and $v$ are  all independent of frequency
and are given by
\begin{eqnarray}
a(\epsilon_{f},\Delta) &=& - \pi \eta_{2}		\\
b(\epsilon_{f},\Delta) &=& \pi^{2}
\left( \frac{1}{2}-q_{0}-\eta_{1} \right)^{2}	\label{def-b}		\\
w(\epsilon_{f},\Delta) &=& \Delta (1-\pi \eta_{2})		\\
v(\epsilon_{f},\Delta) &=&
 - \pi \Delta \eta_{2} \left( \frac{1}{2} - \pi \eta_{2} \right)
- \pi \epsilon_{f} \left( \frac{1}{2}-q_{0}-\eta_{1} \right)	.
	\label{def-v}
\end{eqnarray}
By carrying out integration, we find
\begin{eqnarray}
\Lambda_{2}(\nu_{0}, \eta_{1},\eta_{2},\epsilon_{f},\Delta)
 &=&	\frac{w}{2\pi} \ln\left( \ln^{2}\nu_{0}
+ a \, \ln\nu_{0} + b \right)		\nonumber   \\
&-& \frac{2v-aw}{2\pi}  \times \left\{
\begin{array}{ll}
\frac{1}{\sqrt{a^{2}-4b}} \ln
\left(\frac{ 2 \ln\nu_{0} +\sqrt{a^{2} - 4 b} }
{2 \ln\nu_{0} - \sqrt{a^{2} - 4 b} } \right) ,
&  a^{2}-4b > 0  \\
\frac{2}{\sqrt{4b-a^{2}}} \tan^{-1} \left( \frac{ \sqrt{4b-a^{2}} }
{ 2 \ln\nu_{0} + a }  \right) ,   &  a^{2}-4b < 0
\end{array}		\right.		\label{Lamb2}
\end{eqnarray}

{}From the definition of $\Lambda_{1}$, it is an integral of  the type
\begin{equation}
D \Lambda_{1}(D, \eta_{1},\eta_{2})
 - \nu_{0} \Lambda_{1}(\nu_{0},\eta_{1},\eta_{2})
= \int^{D}_{\nu_{0}} \frac{d\nu}{2\pi}
\ln \left( \ln^{2}\nu + a \ln\nu + b \right) ,
\end{equation}
where we choose $\nu_{0}$ big enough so that the argument
of the log function is always positive.
We can see that $\Lambda_{1}(\nu_{0}, \eta_{1},\eta_{2})$
is analytic in $a$ and $b$ for small values of $a$ and $b$.
In some cases, $\Lambda_{1}$ can be expressed in terms of
the  standard integral of exponential functions such as $Ei(x)$.
In the present problem, the parameters $a$ and $b$ never
get very big. A series expansion is sufficient for the  practical purpose.
The expression we used in the  present calculation is,
\begin{eqnarray}
 \pi \Lambda_{1}(\nu_{0}, \eta_{1},\eta_{2}) &=&
\left[  \frac{P_{1}(\ln^{-1}\nu_{0})}{ \ln\nu_{0} }
- \frac{ Ei(\ln\nu_{0}) }{\nu_{0}} \right]
 \left( e^{\sqrt{a^{2}/4-b}} + e^{-\sqrt{a^{2}/4-b}} \right) e^{-a/2}
		\nonumber 		\\
&+& \sum_{n=1}^{m} (-1)^{n+1} \frac{ P_{n}(\ln^{-1}\nu_{0}) }
{n \ln^{n} \nu_{0} }  ( \alpha_{1}^{n} + \alpha_{2}^{n} )
+  2\left[ \ln\ln\nu_{0}
-  \frac{P_{1}(\ln^{-1}\nu_{0})}{ \ln\nu_{0} }  \right]   ,  \label{Lamb1}
\end{eqnarray}
where $P_{n}$ are polynomials of $\ln^{-1}\nu_{0}$,
\begin{equation}
P_{n}(x)= 1 + n x + n(n+1) x^{2} + \cdots
+   n(n+1)\cdots(m-1)  x^{m-n}		,
\end{equation}
and $\alpha_{1}$, $\alpha_{2}$ are related to $a$, $b$ through
\begin{equation}
\alpha_{1}+\alpha_{2}=a, \hspace{.4in} \alpha_{1}\alpha_{2} =b .
\end{equation}
$Ei(x)$ is the standard integral of exponential function, defined by
\begin{equation}
Ei(x) =  /\!\!\!\!\!\int^{x}_{-\infty} \frac{e^{t}}{t} dt .
\end{equation}
Note that $ \alpha_{1}^{n} + \alpha_{2}^{n} $ are expressed
as polynomials of $a$ and $b$.
 In the expansion (\ref{Lamb1}), $m$ is
the order of expansion.
The neglected  terms are of   the order
 $[{\rm Max}(|\alpha_{1}|, |\alpha_{2}|)/\ln\nu_{0}]^{m+1}/m$.
Typical values used in our calculation are $m \sim 10-15$
and $\ln \nu_{0} \sim 5-8$.

\appendix{High Field Asymptotics of the Magnetization}

The small
field asymptotic behavior of (\ref{give-cn})
 is the well known  result of the present approach \cite{read83}.
Here, we prove the high field asymptotics for $q_{0}=1/2$ and $q_{0}=1/N$.
 The proof for other values of $q_{0}$ goes parallel.
We shall set $T_{K}=1$ and
 omit the star sign in the notation of saddle point solution
$\epsilon_{f}^{*}(h)$ and $\Delta^{*}(h)$.

Let's first consider $q_{0}=1/2$ and even $N$.
In the high magnetic field, the ''$f$-electron'' level is
split into $N$ levels. Each of them is distant from the others.
For $q_{0}=1/2$,  the ''$f$-electrons''
occupy the lowest $N/2$ levels: $\sigma=-S, \;  -S+1, \, \cdots, -1/2$.
The $\sigma=-1/2$ level will  lie  close
to the Fermi level. Spin exchange will result
in a small resonant width.
Thus, we write the solution in the form,
\begin{equation}
\epsilon_{f} =  \frac{h}{2}  - \delta \epsilon_{f},
\hspace{.4in} \frac{ \delta \epsilon_{f} }{h}, \;
\frac{ \Delta }{h} \rightarrow 0 ,
\hspace{.2in} {\rm as}  \hspace{.1in} h \rightarrow \infty  .
\end{equation}
We recall that $S$ is the spin and $N=2S+1$.
Since we are looking for $\ln^{-1}h$ asymptotic terms,
we neglect all terms which die as $h^{-1}$ or faster.
Thus,
\begin{equation}
\epsilon_{f\sigma} = \epsilon_{f} + \sigma h =
\left\{ \begin{array}{ll}
\left(\sigma + \frac{1}{2} \right) h ,
 &  \hspace{.2in} \sigma \neq - \frac{1}{2}	\\
- \delta\epsilon_{f} , & \hspace{.2in}	\sigma = - \frac{1}{2}
	\end{array}		\right.			\label{bigh-simp}
\end{equation}
With this approximation,
the magnetization is simplified to
\begin{equation}
M =M_{0} - \frac{1}{ 2 \pi}
\tan^{-1}\left(\frac{\Delta}{\delta\epsilon_{f}}\right)
-   \int^{D}_{0} \frac{d\nu}{2\pi}
\frac{1}{\Gamma(\nu)} \left[ \Gamma_{\lambda}(\nu)
\frac{\partial \Gamma_{r}}{\partial h}  +
 \Gamma_{r}(\nu)
\frac{\partial \Gamma_{\lambda}}{\partial h}  +
2  \Gamma_{\lambda r}(\nu)
 \frac{\partial \Gamma_{\lambda r}}{\partial h}  \right]	,
		\label{simplif-M}
\end{equation}
where $M_{0}= \sum_{\sigma > 0} \sigma$,  is the saturation
value of the magnetization.
To shorten the notation, we use the unregularized $1/N$
fluctuation energy (\ref{sum-int}) to carry out the proof.
Since the values for $\delta\epsilon_{f}$ and $\Delta$
are given by the saddle point equations~(\ref{sad1-expre})
and (\ref{sad2-expre}), we have to make use of them.
With the simplification~(\ref{bigh-simp}),
The equation~(\ref{sad1-expre}) is similarly reduced to
\begin{equation}
- \frac{1}{\pi}
\tan^{-1}\left(\frac{\Delta}{\delta\epsilon_{f}}\right)
+  \int^{D}_{0} \frac{d\nu}{2\pi}
\frac{1}{\Gamma(\nu)} \left[ \Gamma_{\lambda}(\nu)
\frac{\partial \Gamma_{r}}{\partial \epsilon_{f}}  +
 \Gamma_{r}(\nu)
\frac{\partial \Gamma_{\lambda}}{\partial \epsilon_{f}}  +
2  \Gamma_{\lambda r}(\nu)
 \frac{\partial \Gamma_{\lambda r}}{\partial \epsilon_{f}}  \right] =0
		\label{simplif-sad1}
\end{equation}
The matrix element
$\Gamma$'s involve the spin component summation
$\sum_{\sigma}$,
\[ \Gamma_{\lambda}(\nu) =
\frac{1}{N} \sum_{\sigma} \Gamma_{\lambda}^{(\sigma)}(\nu),
\hspace{.2in}
\Gamma_{r}(\nu) =
\frac{1}{N} \sum_{\sigma} \Gamma_{r}^{(\sigma)}(\nu),
\hspace{.2in}
 \Gamma_{\lambda r}(\nu) =
\frac{1}{N} \sum_{\sigma} \Gamma_{\lambda r}^{(\sigma)}(\nu) \]
Each spin component $\Gamma^{(\sigma)}$
 of the $\Gamma$'s can be read off from
(\ref{Gam-l})-(\ref{Gam-r}).
The difference between the derivatives of the $1/N$ free energy
appearing in (\ref{simplif-M}) and  (\ref{simplif-sad1}) is
that $\partial/\partial h$ in (\ref{simplif-M}) will bring down
an additional  factor $\sigma$ with respect
to $\partial/\partial \epsilon_{f}$.
Dividing (\ref{simplif-sad1}) by two
and subtracting it from (\ref{simplif-M}),
we find
\begin{equation}
 M = M_{0} - \frac{1}{N} \sum_{\sigma}
\left(\sigma + \frac{1}{2} \right)
 \int^{D}_{0} \frac{d\nu}{2\pi \, \Gamma(\nu)} \left[ \Gamma_{\lambda}(\nu)
\frac{\partial \Gamma_{r}^{(\sigma)}}{\partial \epsilon_{f} }  +
 \Gamma_{r}(\nu)
\frac{\partial \Gamma_{\lambda}^{(\sigma)}}{\partial \epsilon_{f} }  +
2  \Gamma_{\lambda r}(\nu)
 \frac{\partial \Gamma_{\lambda r}^{(\sigma)}}{\partial \epsilon_{f} }
 \right]	.
\end{equation}
Note that the $\sigma=-1/2$ component vanishes in the above $\sigma$ summation
so we can replace $\epsilon_{f\sigma}$ by $(\sigma + 1/2 )h$.
Carrying out the derivatives, we find
\[
M = M_{0} - \frac{1}{\pi N} \sum_{\sigma \neq -\frac{1}{2}}
\int_{0}^{D} \frac{d\nu}{\Gamma(\nu)}
 \frac{ h (\sigma+ \frac{1}{2} )^{2} }{ (\sigma+ \frac{1}{2} )^{2}
h^{2} + (\nu+\Delta)^{2} }
   \left\{ \frac{2}{N}
\sum_{\mu=-S}^{S} \ln\left[ \frac{ \epsilon_{f\mu}^{2} +(\nu+\Delta)^{2}}
{ \epsilon_{f\mu}^{2} +\Delta^{2}} \right]
	\right.		\]
\begin{equation}
 +  \left.  \frac{\nu}{N(\nu+\Delta)}  \sum_{\mu}
 \ln \left(  \epsilon_{f\mu}^{2} +\Delta^{2} \right)
+ \frac{4(\nu+\Delta)}{Nh(\sigma+ \frac{1}{2} )}
\sum_{\mu} \left[ \tan^{-1}
\left( \frac{\epsilon_{f\mu}}{\nu+\Delta} \right)
- \tan^{-1} \left( \frac{\epsilon_{f\mu}}{\Delta} \right)	\right]
\right\}	.	\label{M-intermd}
\end{equation}
By noting, from the equation~(\ref{sad2-expre}),
\[  \frac{1}{N} \sum_{\mu=-S}^{S}
\ln \left(  \epsilon_{f\mu}^{2} +\Delta^{2} \right) \sim {\cal O}(1/N) , \]
we can  expand the expression inside curly bracket of
(\ref{M-intermd}) in $1/N$.
We shall also  expand $\Gamma(\nu)$,
\begin{eqnarray}
 \Gamma(\nu) &=& \left[ \frac{1}{N} \sum_{\mu=-S}^{S}
\ln \left(  \epsilon_{f\mu}^{2} +(\nu+\Delta)^{2} \right) \right]^{2}
		\nonumber 		\\
  & & + \left\{ \frac{2}{N} \sum_{\mu=-S}^{S} \left[
\tan^{-1} \left(\frac{\epsilon_{f\mu}}{\nu+\Delta}\right)
-\tan^{-1} \left(\frac{\epsilon_{f\mu}}{\nu+\Delta}\right) \right]
\right\}^{2}	+ {\cal O}(1/N)	.  		\label{Gam-asymp}
\end{eqnarray}
By changing the  dummy variable, $\nu = h \, x$,
we can make following expansion,
\begin{eqnarray*}
  \frac{1}{N} \sum_{\mu=-S}^{S}
\ln \left[  \epsilon_{f\mu}^{2} +(\nu+\Delta)^{2}
\right] &=& 2  \ln h  + \frac{1}{N} \sum_{\mu=-S}^{S}
\ln \left[  (S+\mu)^{2} + x^{2} \right]				\\
 &=& 2 \ln h  \left[ 1+ {\cal O}(\ln x/\ln h)  \right]	,
\end{eqnarray*}
where we dropped  terms of order $\Delta/h$  as usual. That
it is possible to make
$\ln x/\ln h$ expansion in the last expression is due to the
convergence of the integration in (\ref{M-intermd}).
We also expand  $\Gamma(\nu)$, given by (\ref{Gam-asymp}), in $\ln^{-1} h$
and keep the leading term.
The upper integration limit in (\ref{M-intermd})
can be extended  to infinity.
 The final result  for the magnetization is,  after some manipulations,
\begin{eqnarray}
 \frac{M}{M_{0}} &=& 1-  \frac{1}{N M_{0}} \sum_{\sigma \neq -\frac{1}{2} }
\int_{0}^{\infty} \frac{dx}{\pi}
\frac{ \left(\sigma+ \frac{1}{2}\right)^{2} }
{ \left(\sigma+\frac{1}{2}\right)^{2} + x^{2} }
\; \frac{1}{\ln h}
\left[ 1 + {\cal O}(\ln x / \ln h) \right] 	\nonumber 	\\
&=& 1 -  \frac{1}{\ln h}
\frac{1}{2  N M_{0}  } \sum_{\sigma \neq -\frac{1}{2}}
 |  \sigma + \frac{1}{2} |
		\nonumber 		\\
&=& 1 -  \frac{1}{ N \ln h}   + {\cal O}(\ln^{-2}h)   .
\end{eqnarray}

For $q_{0}=1/N$,
strictly speaking, $1/N$ is no longer the loop expansion
parameter. Nevertheless, if we repeat the above steps,
we find
\begin{equation}
\frac{M}{S} = 1  -  \frac{1}{ 2 \ln h}   + {\cal O}(\ln^{-2}h) .
\end{equation}
Note that the leading log correction is independent of $N$ for $q_{0}=1/N$.
It is easy to see this from the perturbation in $g$. This term comes
from the linear term, $g/2$, in the $g \ll 1$ perturbation.
The diagram for this term involves one conduction electron loop
and one ''$f$-electron'' loop which together contribute a factor $N^{2}$.
The interaction vertex brings in a factor $1/N$. After normalization,
{\em i.e.} dividing by $S \sim N$, it is independent of $N$.

\bibliography{hfmag,hightc}


\begin{table}
\caption{Definition of symbols and notations}
\begin{tabular}{cccc}
 Symbol  & Definition (Eq. No.) &  Symbol & Definition (Eq. No.) \\
\tableline
 $D$                 &  Bandwidth	  &
		$\Gamma$	 & (\ref{Gam-exp})	\\
 $\rho(\epsilon)$    &  Density of states &
		$\Gamma_{1}$	 &  (\ref{Gam-1})	\\
 $h$	&	(\ref{coqb})		&
		$\Gamma_{2}$	 &  (\ref{Gam-2})	\\
  $q_{0}$            & (\ref{constraint}) &
		$\eta_{1}$	 &  (\ref{def-eta1})    \\
  $\epsilon_{f}$, $r_{0}$  & (\ref{sad-expan}) &
		$\eta_{2}$	 &  (\ref{def-eta2})  \\
 $\Gamma_{\lambda}$  &	(\ref{Gam-l})  &
		$\Lambda_{1}$	& (\ref{L1}), (\ref{Lamb1})  \\
 $ \Gamma_{\lambda r}$ & (\ref{Gam-lr})  &
	$\Lambda_{2}$	& (\ref{L1L2}),  (\ref{L2}), (\ref{Lamb2})  \\
 $ \Gamma_{r}$ & (\ref{Gam-r})  &
		$\nu_{0}$	& (\ref{F1-sep})	\\
	 $ g$, $T_{K}^{(0)}$ &  (\ref{Tk0})   &
		$T_{K}$	& (\ref{kondo-T})		\\
  $\epsilon_{f\sigma}$, $\Delta$ & (\ref{def-Delta})	&
	$\widetilde{F}$  & (\ref{def-F-reg}), (\ref{F-til-exp})	\\
 $ F_{\rm MF}$ 	 & (\ref{f-ener-1}), (\ref{F-mf}) &
  $F_{\rm reg}$ 	& (\ref{def-F-reg}),  (\ref{Freg-fin})	\\
 $ F_{1/N}$	& (\ref{f-ener-1}), (\ref{F1}) &
	$F^{\rm reg}_{1/N}$	& (\ref{F-final})	\\
\end{tabular}
\label{symbol}
\end{table}

\begin{table}
\caption{The  calculated magnetic Wilson crossover numbers
 for the Coqblin-Schrieffer model,
$q_{0}=1/N$,  defined as $\alpha'$
of (\ref{give-cn}). With $T_{K}$ defined by (\ref{kondo-T}),
we read off the initial gradient,
$\alpha$ in (\ref{m-h-expre}),  the magnetization curve.
Then the crossover number is $\alpha'=\alpha/(1+2\ln 2/N)$. }
\begin{tabular}{ccc}
 N 	& 	Crossover number & Bethe Ansatz 	\\
\tableline
 2 	&	0.25	&	0.342  (=$1/\sqrt{e \pi}$)  	\\
 4	&	0.65	&	-				\\
 6	&	1.01	&	-				\\
 8	&	1.36	&	-				\\
 10	&	1.70	&	-				\\
\end{tabular}
\label{c-number}
\end{table}

\figure{The solution of the  saddle point equations
in the magnetic field for $q_{0}=1/6$ and $N=6$.
$T_{K}$ is defined by (\ref{kondo-T}).  $\epsilon_{f}$ is the position
of the resonant level and $\Delta$ is the width.
	\label{fig-sadd}	}

\figure{The universal magnetic field dependent magnetization
for $q_{0}=1/2$ and  for $N=2$(short dashed line),
 $N=4$(long dashed line),  $N=6$(dash-dotted line),
 $N=8$(solid line). All curves are parameter free.
Note the improving quality for larger $N$.
	\label{fig-qhalf} }

\figure{The universal magnetic field dependent magnetization
for the Coqblin-Schrieffer model, {\em i.e.} $q_{0}=1/N$, and for
$N=6$(dashed line),  $N=8$(dash-dotted line),
 $N=10$(solid line). All curves are parameter free.
The points are Hewson and Rasul's Bethe Ansatz
results: $N=6$(filled triangles), $N=8$(filled circles).
The proportionality factor between $T_{K}$
defined by (\ref{kondo-T}) and the $T_{1}$ appearing in Bethe
Ansatz solution is determined for each $N$ by matching the
small field gradient of the magnetization.
\label{fig-coqb}	}

\end{document}